\documentclass{svproc}

\usepackage{indentfirst, times, graphicx, subfigure,
setspace, amsmath, multirow, epstopdf, amssymb}

\usepackage{bbding}

\usepackage{array}
\usepackage{rotating}
\usepackage{pdflscape}
\usepackage[sort]{natbib}
\usepackage{algorithm}
\usepackage{algpseudocode}

\usepackage{lmodern} 
\usepackage[T1]{fontenc} 
\usepackage[nottoc]{tocbibind} 
\usepackage[pdfborder={0 0 0}]{hyperref}
\usepackage[all]{hypcap}

\graphicspath{{figures/}}

\usepackage{url}

\begin{document}
\mainmatter              
\title{HSEARCH: fast and accurate protein sequence motif search and clustering}
\titlerunning{HSEARCH: protein motif search and clustering}  
%
\author{Haifeng Chen \and Ting Chen \Envelope}
\authorrunning{H. Chen and T. Chen} 
%
\tocauthor{Haifeng Chen, and Ting Chen}
\institute{Molecular and Computational Biology, \\ University of Southern California, Los Angeles, 90089, USA \\
\email{tingchen@usc.edu}}

\maketitle              

\setlength{\baselineskip}{1.3em}

\begin{abstract}
Protein motifs are conserved fragments occurred frequently in protein sequences.
They have significant functions, such as active site of an enzyme.
Search and clustering protein sequence motifs are computational intensive. 
Most existing methods are not fast enough to analyze large data sets for motif finding
or achieve low accuracy for motif clustering.
We present a new protein sequence motif finding and clustering algorithm, called HSEARCH.
It converts fixed length protein sequences to data points in high dimensional space, 
and applies locality-sensitive hashing to fast search homologous protein sequences for a motif.
HSEARCH is significantly faster than the brute force algorithm  for protein motif finding and 
achieves high accuracy for protein motif clustering.
\end{abstract}

\section{Introduction}
Protein sequence motifs are conserved fragments 
occurred frequently in protein sequences.
Protein motifs have significant functions,  
such as active site of an enzyme \citep{grant2011fimo}.
A protein sequence motif is a set of same-length 
homologous protein sequences.
A protein sequence motif could be represented by 
a position weight matrix (PWM) \citep{stormo1982use}.
Each element in PWM represents the percentage
of an amino acid in a  specific position.
For any protein sequence,  the probability that it 
belongs to a motif could be calculated using the PWM.

Several motif finding algorithms have been developed in the past decades, 
such as MotifScanner \citep{mele2016arabidopsis},
MotifViz \citep{fu2004motifviz},
STORM \citep{schones2007statistical}, 
PoSSuMsearch \citep{beckstette2006fast} 
and FIMO \citep{grant2011fimo}.
PoSSuMsearch and FIMO support protein motif finding, 
and  others only support DNA motif finding.
They are web browser-based programs which cannot
analyze a huge amount of protein sequences.
Nowadays, more and more protein sequences have been 
produced by next-generation sequencing machines,
especially in metagenomic studies. For example, 
IGC data set in human gut microbiome \citep{qin2010human} 
has about 10 million protein sequences with total length 2.4 billion amino acids.
It is not computational practical to search these protein sequences on web browser-based programs.

Additionally, 
sequence clustering is a fundamental technique which has many important applications.
clustering $k$-mers from a large protein database could discover new protein motifs.
Several sequence clustering programs have been developed in the past decades. 
For example, CD-HIT \citep{li2006cd} uses incremental greedy method. 
The first sequence is set as a representative sequence for the first cluster.
Then each query sequence is compared with representative sequences in existing clusters.
If the similarity between the query sequence and 
the current compared representative sequence is higher than certain threshold, 
the query sequence is added to that cluster.
If no existing representative sequence is found, a new cluster is
generated and the query sequence is the representative sequence for the new cluster. 
UCLUST \citep{edgar2010search} is similar to CD-HIT but uses different
lengths of $k$-mers to filter false positive candidates.
kClust \citep{hauser2013kclust} improves the accuracy
by sorting all sequences and searching all the existing representative sequences.  
However,  these existing clustering algorithms 
have low accuracy in clustering protein sequences since 
they cluster protein sequences based on sequence identify,
but most homologous protein sequence have low sequence identify.
It is better to cluster protein sequence based on sequence similarity rather than 
sequence identity.

Here we present a new protein sequence motif finding and clustering algorithm, called HSEARCH.
HSEARCH converts fixed length protein sequences to data points in high dimensional space, 
and applies locality-sensitive hashing to fast search homologous  protein sequences for a motif .
HSEARCH is significantly faster than the brute force algorithm  for protein motif finding and 
achieves high accuracy for protein motif clustering.

\section{Methods}
HSEARCH	 converts proteins sequences and protein motifs to data points in high dimensional space 
and applies fast neighbor search method locality-sensitive hashing (LSH) to finding near data points for 
a protein motif.  To reduce distortion of the conversion between protein sequence and data point,
if two protein sequences have high similarity, the converted data points for them could have small distance.
Similarly, for two sequence have low similarity,   the converted data points should have large distance.
Meanwhile,  all converted data points should be in metric space in order to apply LSH technique.

\subsection{Similarity matrix to distance matrix}
The similarity score of two protein sequences is the sum of scores for every pair of amino acids.  
The similarity score of each pair of amino acids usually are defined in BLOSUM62 matrix \citep{henikoff1992amino}.
To obtain the converted coordinate of each amino acid, BLOSUM62 similarity score matrix should be 
converted to distance matrix and the convention  satisfies four conditions shown in \textit{Definition 1}.

\vspace{1em}
\label{conditions-matrix}
\noindent\textbf{\textit{Definition 1} } 
Let $s(i, j)$ denote the similarity score between $i^{th}$ and $j^{th}$ amino acids in BLOSUM 62 matrix
and $d(i, j)$ denotes the distance between converted coordinates for $i^{th}$ and $j^{th}$ amino acids.
Then, for any $ 1 \leq i,j,k,u,v \leq 20$,
\begin{enumerate} 
\item $s(i, j)  \geq s(u, v) \Leftrightarrow d(i, j)  \leq d(u, v)$
\item $d(i, j) = 0 \Leftrightarrow i = j$
\item $d(i, j)  = d(j, i)$
\item $d(i, j)  \leq d(i, k) + d(k, j)$
\end{enumerate}

In HSEARCH, Equation~\ref{eq:BLOSUM62-distance-matrix} was applied to convert BLOSUM62 similarity matrix to distance matrix.
For any $ 1 \leq i,j \leq 20$,
\begin{equation}\label{eq:BLOSUM62-distance-matrix}
   d(i, j) = s(i, i) + s(j, j) - 2\times s(i, j) 
\end{equation}

Since BLOSUM62 similarity matrix  is symmetric, 
it is easy to prove that Conditions 2 and 3 hold in  \textit{Definition 1}.
We manually validated 8,000 different combinations of triangle inequality for Condition 4, 
fortunately,  Equation~\ref{eq:BLOSUM62-distance-matrix} satisfies all 8,000 combinations for BLOSUM62 matrix.

\begin{figure}[ht]
\centering
\includegraphics[width=0.6\textwidth]{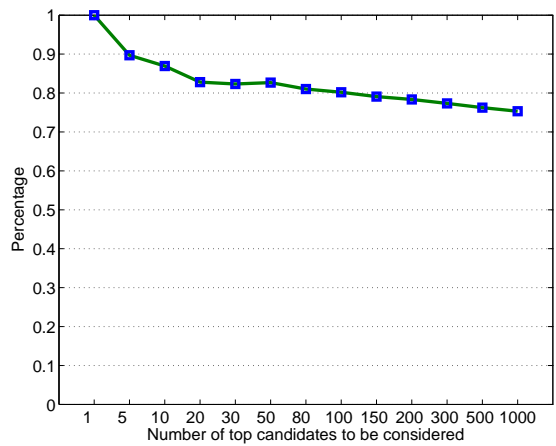}
\caption{Percentage of $25$-mers conserved during the conversion of similarity matrix to distance matrix}
\label{fig:distortion_conversion}
\end{figure}

Condition 1 in \textit{Definition 1} is not fully satisfied.
We randomly generated 100,000 $25$-mers to estimate the distortion during this conversion.
For each $25$-mer, we used both BLOSUM62 similarity matrix and distance matrix to obtain
top $q$ most similar or close $25$-mers, respectively.
We got the percentage of candidates by using BLOSUM62 and also in the candidates by using distance matrix.
As shown in Figure~\ref{fig:distortion_conversion}, when $q$ increases, the percentage decreases a little bit,
but not much. More than 80\% of 25-mers are conserved during this conversion, which is good enough since
BLOSUM62 similarity matrix itself is an estimation.

\subsection{Distance matrix to coordinates}
Multidimensional scaling (MDS)  is the technique to convert distance matrix to coordinates
\citep{borg2005modern, groenen2005multidimensional, kruskal1978multidimensional}.

\vspace{1em}
\noindent\textbf{\textit{Definition 2} (Multidimensional Scaling) }
Given a distance matrix  \break $D=[d(i ,j)]$, $(1 \leq i,j \leq 20)$,
find $x_{1}, x_{2}, x_{3},\dots, x_{20}\in\Re^d$, subject that 
\begin{equation}\label{eq:mds}
\sigma^2(x_{1}, x_{2}, x_{3},\dots, x_{20})=\sum_{1 \leq i<j \leq 20}  (d(i, j) - \Vert x_{i}-x_{j}\Vert)^2
\end{equation}
is minimized.
This is also refereed as the least-squares MDS model \citep{groenen2005multidimensional}.
$\Vert x_{i}-x_{j}\Vert$ denotes the Euclidean distance between data point $x_i$ and $x_j$.

The least-squares MDS cannot be solved in closed form and it was 
solved by iterative numerical algorithm, such as SMACOF algorithm 
\citep{de1984upper, de1980multidimensional, de1988convergence, de2005applications}. 
In our experiment, we used Matlab \textit{mdscale} function to obtain $x_{1}, x_{2}, x_{3},\dots, x_{20}$.

\begin{figure}[ht]
\centering
\includegraphics[width=0.6\textwidth]{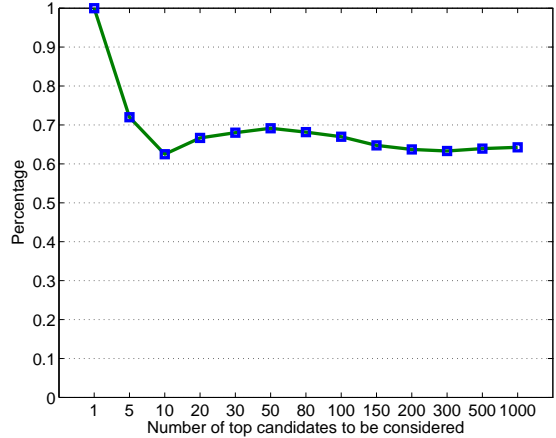}
\caption{Percentage of $25$-mers conserved during the conversion of similarity matrix to coordinates}
\label{fig:distortion_conversion_2}
\end{figure}

The coordinate for a protein sequence is the concatenation of coordinates for all amino acids. 
We randomly generated 100,000 $25$-mers to estimate the distortion for conversion from 
similarity matrix to coordinates.
For each $25$-mer, we used both BLOSUM62 similarity matrix and coordinates to obtain
top $q$ most similar or close $25$-mers, respectively.
As shown in Figure~\ref{fig:distortion_conversion_2}, when $q$ increases, the percentage decreases a little bit,
but more than 60\% of 25-mers are conserved during this conversion.


\subsection{Locality-Sensitive Hashing} \label{sec:lsh}
Locality-Sensitive Hashing (LSH) is  one of the popular used methods for searching near neighbors. 
It finds all data points $p_{1}, p_{2}, p_{3},\dots, p_{n}$, whose distance is less than a threshold $T$ to a given data point $q$
  $\in\Re^d$ \citep{datar2004locality, har2012approximate, Andoni2004}.
LSH functions are designed that data points closer to each other have more chance to 
be hashed into the same bucket than data points far from each other, which is demonstrated in Equation~\ref{eq:lsh1}.
Here $H$ denotes a LSH function which hashes a data point to an integer, and $Pr$ denotes the probability.
\begin{equation}\label{eq:lsh1}
Pr[H(p_{i}) = H(p_{j})] > Pr[H(p_{u}) = H(p_{v})]  \Leftrightarrow \Vert p_{i}-p_{j}\Vert < \Vert p_{u}-p_{v}\Vert
\end{equation}

LSH originally is designed for Hamming Distance in binary data, 
and later \citet{datar2004locality} extended LSH to support for 
Euclidean Distance in high dimensional data points. 
LSH function for Euclidean Distance is based on random projection.
Two data points are closer in high dimensional space intuitively 
 that they could be projected near to each other in a random line.
 If the random line is chopped to buckets with length $w$, two points
 with smaller distance have higher probability to project into the
 same bucket.
\begin{equation}\label{eq:lsh_projection_1}
h_{a,b}(p) = \lfloor \frac{a \cdot p + b}{w}\rfloor
\end{equation}

Equation~\ref{eq:lsh_projection_1} shows the LSH function family for any data point $p \in \Re^d$.
$a$ is a $d$-dimensional random vector from Gaussian distribution, and $b$ is a real number
uniformly randomly drawn from $[0, w)$, where $w$ is the bucket width.  
The probability of two data points with distance $c$  projected into the same bucket is 
shown in Equation~\ref{bucket_probability_1} (More details in Appendix~\ref{App:AppendixA}).
\begin{equation} \label{bucket_probability_1}
   Pr[c]    =   \int_{0}^{w}\frac{1}{c} f_{Y}(\frac{t}{c})   (1-\frac{t}{w})dt \\
\end{equation}

\noindent $f_{Y}(y)$ is the probability density function for half-normal distribution. 
$Pr[c]$  is monotonically decreasing in $c$ when the bucket width $w$ is fixed, which
satisfies the propriety shown in Equation~\ref{eq:lsh1}

In order to reduce the chance that two points far from each other hash to the same bucket,
normally, $K$ random lines are selected. If two data points are projected to the same
bucket in all $K$ random lines,  the two data points are stored in the same bucket in a 
hash table. Additionally, to increase the chance that two points near to each other
 hash to the same bucket, generally, $L$ hash tables are built on the data set.
Thus, for any data point $p$, the hash value $h(p)$ is shown as follow:
\begin{equation} 
h(p)=(h_{a_{1},b_{1}}(p), h_{a_{2},b_{2}}(p) \dots, h_{a_{K},b_{K}}(p)) \\
\end{equation}

To search neighbors for a data point $q$, all data points in the same hash bucket with $q$ in 
one of the hash tables will be validated.  The probability that two points with distance $c$
hash to the same bucket in of the of $L$ hash table is 
\begin{equation}
Pr_{K,L}[c] = 1- (1- Pr[c]^K)^L
\end{equation}

\noindent The increasing of $K$ could decrease the probability, and the
increasing of $L$ could increase the probability.

\section{Results}

\subsection{Protein Motif Finding}

For a given protein database, each protein sequence is converted to $l-k+1$ $k$-mers
where $l$ is the length of protein sequence.
A $k$-mer is converted to a $d$-dimensional data point. 
LSH tables are built on all data points from the protein database.
A motif is also converted to a data point. The coordinate of a motif is obtained 
by the concatenation of coordinate in each position. 
The coordinate of a motif is also refer as the center point of the motif.
The coordinate for each position $j$ ($1 \leq j \leq l$)
is shown in Equation~\ref{coordinate_motif}. $M_{ij}$ is the 
element in PWM which represents the probability of $i^{th}$ amino acid at $j^{th}$ position, 
and $x_i$ is the coordinate for $i^{th}$ amino acid.

\begin{equation} \label{coordinate_motif}
C_j = \sum_{i=1}^{20} (x_i \times M_{ij})
\end{equation}

The query $q$ represents the motif and applies  LSH technique 
to search all data points $p$ in the database where $\Vert p - q \Vert \leq T$.

\begin{figure}[ht]
\centering
\includegraphics[width=0.8\textwidth]{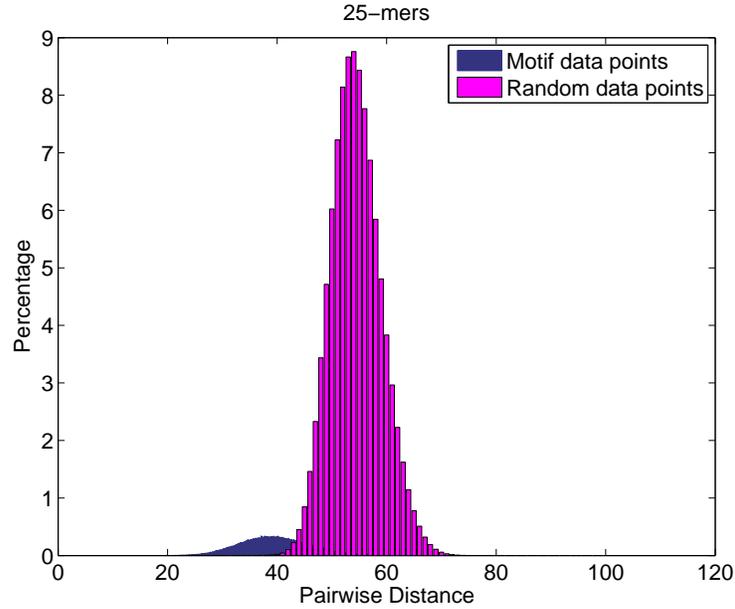}
\caption{Percentage of data points inside and outside motifs to the centers as a function of distance}
\label{fig:distance2centers}
\end{figure}

\begin{figure}[ht]
\centering
\includegraphics[width=0.7\textwidth]{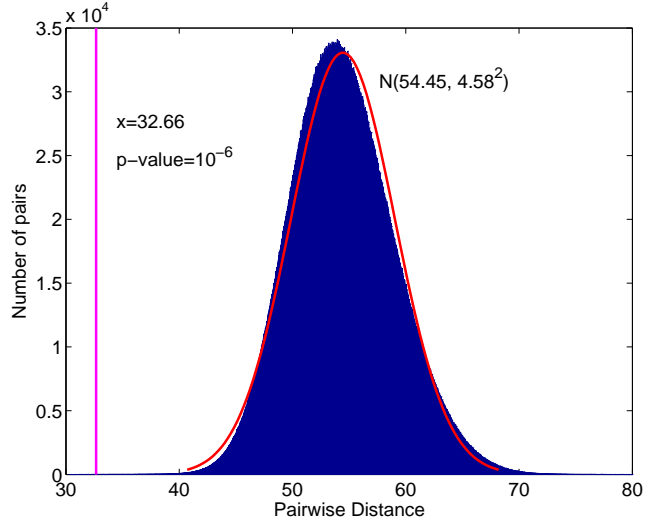}
\caption{The number of data point pairs as a function of distance}
\label{fig:random_pairwisedistance}
\end{figure}

Figure ~\ref{fig:distance2centers} shows the pairwise distance among sequences in a motif
to the center point from data points in motifs and random data points. 
The data points in a motif and random data points are well separated.
In order to set the threshold $T$, 
we randomly selected 100,000 $25$-mers from IGC protein database, 
 and plotted the frequency of pairwise distance from these $25$-mers to centers of  motifs,
  as shown in Figure~\ref{fig:random_pairwisedistance}. 
 The frequency roughly follows a normal distribution $N(54.45, 4.58^2)$, 
 and the probability that pairwise distance  less than 32.66 is $1.0\times10^{-6}$. 
 Thus, if a data points in LSH table has a distance less than 32.66 to the query, 
 the corresponding $25$-mer belongs to the motif represented by the query with  $p$-value $1.0\times10^{-6}$. 
 Therefore, for fixed length motif  sequences, we set the threshold $T$ to 32.66 for $25$-mers. 
 In our results, we also tested  $p$-value $1.0\times10^{-4}$ with threshold $T$ 34.90.
 Different lengths of $k$-mers have different thresholds which are obtained from Pfam database.

\noindent \textbf{Data sets}
We compared the performance of HSEARCH with brute force algorithm on a data set
with 18,000,000 $k$-mers.
The length of $k$-mers is 25, and all of them are randomly selected from IGC data set \citep{qin2010human}. 
Furthermore, we randomly selected 574 motifs from Pfam-A.seed database \citep{bateman2004pfam}, 
and each motif with at least 50 protein sequences. All motif sequences are trimmed 
to 25-mers and converted to $d$-dimensional data points. 
The center of each motif sequences are treated as queries for 
searching $k$-mers from IGC data set,
and a $k$-mer with a distance less than threshold $T$ to the center point of a motif, then the $k$-mer is 
identified as a sequence of that motif.

Let the set $P=\{p_1, p_2, p_3, \dots, p_n\}$ are the  $k$-mers and $C=\{c_1,c_2, c_3, \dots, c_m\}$ are the
centers of motifs. The result of the brute force algorithm is a set of pairs 
$R_{b}=\{(p_i, c_j) | \Vert p_i - c_j \Vert \leq T, 1\leq i\leq n, 1 \leq j \leq m \}$ and the result of HSEARCH $R_H$
is  a subset of $R_b$.
\begin{equation}\label{eq:weight}
weight(p_i, c_j) = 
\begin{cases}
\frac{1}{\Vert p_i - c_j  \Vert - \lambda} &  \Vert p_i - c_j  \Vert - \lambda > 1.0 \\
1.0 & else
\end{cases} 
\end{equation}
\begin{equation}\label{eq:tp2}
TP= \sum_{i=0}^{n} \sum_{j=0}^{m} weight(p_i, c_j) I[(p_i, c_j) \in R_b \land (p_i, c_j) \in R_H]
\end{equation}
\begin{equation}\label{eq:fn2}
FN= \sum_{i=0}^{n} \sum_{j=0}^{m} weight(p_i, c_j) I[(p_i, c_j) \in R_b \land (p_i, c_j)  \notin R_H]
\end{equation}
\begin{equation}\label{eq:recall2}
Recall = \frac{TP}{TP + FN}
\end{equation}

Brute force algorithm obtains all pairs of $(p_i, c_j)$, where $\Vert p_i - c_j  \Vert \leq T$. 
We access the accuracy of HSEARCH using recall defined in Equation~\ref{eq:recall2}.
True positive (TP) and False negative (FN) are defined in Equations~\ref{eq:tp2} and~\ref{eq:fn2}.
Each pair has a  weight which is defined as shown in Equation~\ref{eq:weight}, 
where  $\lambda$ is the minimal distance among all pairs.
\begin{table}[ht]
\caption{Runtime(hours) for HSEARCH on different percentages of brute force result  }\label{tab:runtime_hsearch1}
\centering 
\def\arraystretch{1.2}%
\begin{tabular}{|c|c|c|c|c|c|c|c|c|c|c|c|} \hline
$p$-value  &    &  10\%	& 20\%  & 	30\%  & 40\%  & 	50\%	 & 60\% & 70\%  & 80\%  & 90\% &100\% \\\hline
\multirow{2}{*}{\footnotesize{$10^{-6}$}} &Runtime     & 0.07	&	0.10 &	0.15 &	 0.19 & 0.23 &	0.37  &	0.53  & 0.60  & 0.78 &  0.92\\\cline{2-12}
                                                                                 &Speedup     & 24.6	&  17.2	 &	11.5 &  9.1   & 7.5	 &	4.6	  & 3.2   & 2.9   &   2.2  &  1.9      \\\hline
\multirow{2}{*}{\footnotesize{$10^{-4}$}} &Runtime     & 0.07	&	0.10 &	0.15 &	0.20  & 0.26 &	0.46  &	0.54  &  0.58 & 1.05 &  1.11 \\\cline{2-12}
 																					   &Speedup     & 24.6	& 17.2	 &	11.5 &	8.6	  &	6.6	 &	3.7	  &	3.2	  &	3.0	  &	1.6	 &	1.5     \\\hline
\end{tabular}
\end{table}

\noindent \textbf{Runtime and Accuracy}
 We assessed HSEARCH runtime on different percentages (Recall defined in Equation~\ref{eq:recall2}) of brute force results.
 Brute force algorithm took 1.72 hours for searching 18,000,000 $k$-mers on 574 motifs.
 Table~\ref{tab:runtime_hsearch1}  shows the runtime on
 different percentages of brute force
 results for $p$-value $1.0\times10^{-6}$ and $1.0\times10^{-4}$. 
 With fixed $K$ and $L$, we used different
 bucket width $w$ in the LSH method to obtain variety of percentage of brute force results. 
 
 As  table shown,
 HSEARCH is about 1.5$\times$  faster than brute force method when getting the same result.
 With  decreasing of the percentage, the speedup of HSEARCH increased significantly. 
 Importantly, since in LSH method, the data points closer to the query has higher probability 
 to be located in the same hash bucket, with  decreasing of the percentage, most of 
 data points with larger distance to the query could be excluded in HSEARCH result, and the 
 data points near the query are still in the HSEARCH result.

\subsection{Motif Sequence Clustering}

HSEARCH uses an incremental greedy method to partition
protein sequences $S=\{s_1, s_2, s_3, \dots, s_n \}$ into a set of clusters
 $C=\{c_{1}, c_{2}, c_{3},\dots,c_{m}\}$.
For a protein sequence $s_{i}$, which has not been added to any cluster,
it is set as the representative sequence of a new cluster $c_{j}$. 
Protein sequences that have not been added to any cluster and
whose distance to $s_{i}$
is less than a threshold $T$ are added to $c_{j}$.
Locality-sensitive hashing technique is applied to search 
 near neighbors for each representative sequence.
 
 \begin{figure}[ht]
\centering
\includegraphics[width=0.7\textwidth]{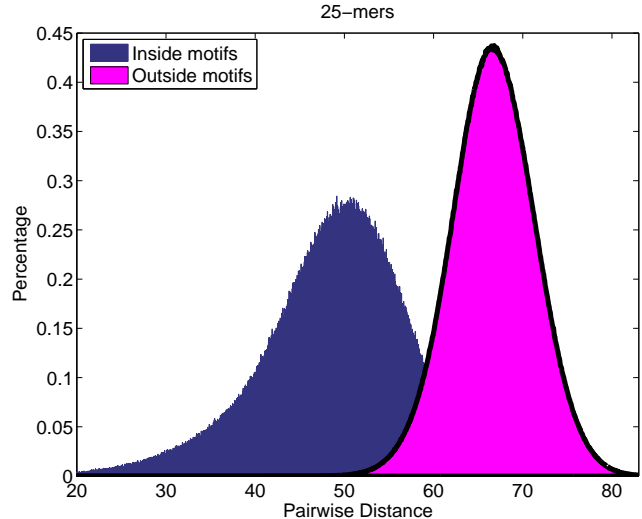}
\caption{Pairwise distance inside and outside motifs for 25-mers}
\label{fig:pairwise_distance_motifs-25mers}
\end{figure}

\noindent \textbf{Data sets}
We assessed the performance of HSERACH for clustering protein sequences on a ground truth data set
selected from Pfam database. 
This data set contains 1,000 motifs with total 316,097 protein sequences.
All protein sequences are trimmed to length 25.
HSEARCH is compared with CD-HIT, UCLUST and kClust.
If a pair of protein sequences are from one moitf, and a program clustered them to the same cluster,
this pair is counted as true positive (TP), otherwise false negative (FN).
Similarly, if a pair of protein sequences are from different motifs,
and a program clustered them to the same cluster, this pair is counted as false positive (FP),
otherwise true negative (TN). 
The recall, precision and F1 score defined in Equations~\ref{eq:recall}, ~\ref{eq:precision}
and~\ref{eq:F1} are used to evaluate the results.
\begin{table}[ht]
\caption{Runtime  (hours), Recall, Precision, F1 and NIM for CD-HIT, UCLUST and kClust in motif clustering}\label{tab:accuracy-clustering}
\centering 
\def\arraystretch{1.1}%
\begin{tabular}{|c|c|c|c|c|c|c|c|c|c|c|}
\hline
\multicolumn{2}{|c|}{ Similarity} & 10\%	& 20\%  & 	30\%  & 40\%  & 	50\%	 & 60\% & 70\%  & 80\%  & 90\%  \\\hline
\multirow{5}{*}{\footnotesize{CD-HIT}} &	Runtime  & 	-&    -   &   - &  -&    -  &   -  &  0.002 &	0.002&	0.002\\\cline{2-11}
																&	Recall  & 	-&    -   &   - &  -&    -  &   -  &  0.025 &	0.011&	0.002\\\cline{2-11}
                                                           &  Precision     & -	& -      &   - 	 &  	-&    -     &    -  &   0.998&	0.999&	1.000\\\cline{2-11}
                                                           &   \textbf{F1}    & -	& -      &    	- &  -	&    -     &  -    &  \textbf{0.049}&	 \textbf{0.021	}& \textbf{0.004} \\\cline{2-11}
                                                           & NMI    & -	& -      &    	- &  -	&    -     &  -    & 0.752 & 0.731 &	0.710 \\\hline                    
\multirow{5}{*}{\footnotesize{UCLUST}} &  Runtime& 0.005 &	0.005	&0.005	&0.005	&0.012	&0.015	&0.010	&0.008&	0.007 \\\cline{2-11}
                                                             &  Recall&0.092&	0.092&	0.092&	0.091&	0.080&	0.055&	0.023&	0.007&	0.002\\\cline{2-11}
                                                             &  Precision&	0.960&	0.960&	0.960&	0.965&	0.985&	0.997&	0.998&	0.999&	0.999\\\cline{2-11}
                                                             & \textbf{F1}&	 \textbf{0.168}&	 \textbf{0.168}&	 \textbf{0.168}&	 \textbf{0.166}&	 \textbf{0.148}&	 \textbf{0.104}&	 \textbf{0.044}&	 \textbf{0.014}&	 \textbf{0.004 }\\\cline{2-11}
                                                             &NMI&   0.799&	0.799&	0.799&	0.798&	0.793&	0.778&	0.751&	0.725&	0.710\\\hline
 \multirow{5}{*}{\footnotesize{kClust}}&Runtime&-& 0.027&	0.022	& 0.015	&0.012&	0.009&	0.007	&0.012&	0.016 \\\cline{2-11}   
 															  & Recall &-& 0.075	&0.074&	0.071	&0.068	&0.060	&0.029	&0.011&	0.002	 \\\cline{2-11}
                                                          & Precision &-&	0.998&	0.998&	0.998&	0.998&	0.997&	0.997&	0.999	&0.999\\\cline{2-11}	 
                                                          &  \textbf{F1}&-&	 \textbf{0.140	}& \textbf{0.137	}& \textbf{0.133}	& \textbf{0.128}&	 \textbf{0.114	}& \textbf{0.056}	& \textbf{0.022}&	 \textbf{0.004} \\\cline{2-11}
                                                          &NMI&-&0.792&	0.791	&0.789	&0.787	&0.780&	0.753	&0.730	&0.708 \\\hline                
\end{tabular}
\end{table}

\noindent \textbf{Runtime  and Accuracy} We investigated the ground truth data set by plotting  pairwise
distances for data points inside motifs, and between motifs,
as shown in Figure~\ref{fig:pairwise_distance_motifs-25mers}.
Since HSEARCH is a heuristic algorithm, if two data points have distance to the center less than $T$,
it dose not guarantee that the distance between these two data points is less than $T$. 
Therefore, we did not set the threshold $T$ to the intersection
of these two plots in Figure~\ref{fig:pairwise_distance_motifs-25mers}, but a little smaller than the intersection.
In HSEARCH, for 25-mers, the threshold $T$ is set to 50.

\begin{table}[ht]
\caption{Runtime  (hours), Recall, Precision, F1 and NIM for HSEARCH in motif clustering}\label{tab:runtime-clustering-hsearch}
\centering 
\def\arraystretch{1.1}%
\begin{tabular}{|c|c|c|c|}\hline
$L$       & 4	&  8  &  16     \\\hline
Runtime   & 0.032&    0.036  &   0.030 	     \\\hline        	                    
Rcall     &	 0.286 &	0.399  &  0.496  \\\hline
Precision &  0.827 &    0.823  &  0.815	 \\\hline	 
 \textbf{F1}       &   \textbf{0.425} &	 \textbf{0.537}  &   \textbf{0.617}	 \\\hline	 
NMI       &  0.859 &    0.880  &  0.897  \\\hline	
\end{tabular}
\end{table}

Table~\ref{tab:accuracy-clustering}
shows the comparisons of runtime and accuracy for CD-HIT, UCLUST and kClust on different sequence
 identities. CD-HIT does not support sequence identity less than 70\% for 25-mers.
kClust does not support sequence identity less than 20\%. 
Table~\ref{tab:runtime-clustering-hsearch} shows the runtime and accuracy for HSEARCH in motif clustering.
All four programs ran very fast in the ground truth data set.
However, CD-HIT, UCLUST and kClust
have recall less than 10\% for all different sequence identities. As we talked above,
clustering protein sequences based on sequence identity is not sensitive enough, since
two protein sequences may be homologous even they have low sequence identity.
The NMI score (Equation~\ref{eq:NMI}) for HSEARCH is also significant higher than CD-HIT, UCLUST and kClust.

\section{Conclusion}
Massive of protein sequence data have been produced with the advanced sequence technologies.
Fast and accurate analyze these protein sequence data becomes a big need in recent years.
Protein motifs are functional units for proteins and it is important to find  protein motifs
in the large number of protein sequences for further data analysis.
The existing protein motif finding methods are based on brute force algorithm which is 
not fast enough to deal with large protein database.
Additionally, the existing protein sequence clustering algorithm are based on sequence identity
which is not sensitive enough for clustering protein sequence.
HSEARCH converts fixed length protein sequences to data points in high dimensional space, 
and applies locality-sensitive hashing to fast search homologous protein sequences for a motif .
HSEARCH is significantly faster than the brute force algorithm since data points that are close to 
the motif center point have more chance to hash to the same bucket by locality-sensitive hashing.
Data points far from the motif center point have less chance to hash to the same bucket which 
efficiently reduces the number of false positive candidates.
HSEARCH clusters protein sequences based on sequence similarity rather than sequence identify
which achieves both high recall and precision for clustering protein motif sequences.
In future research, HSEARCH could use to cluster large protein database and detect statistical 
significant clusters for discovering new protein motifs.

\bibliographystyle{namedplus}
\bibliography{main}

\clearpage

\begin{center}
\textbf{\huge{Appendix}}
\end{center}

\vspace{1em}
\appendix
\section{Probability of hashing to the same bucket in LSH } \label{App:AppendixA}
LSH originally is designed for Hamming Distance in binary data, 
and later \citet{datar2004locality} extended LSH to support for 
Euclidean Distance in high dimensional data points. 
LSH function for Euclidean Distance is based on random projection.
Two data points are closer in high dimensional space intuitively 
 that they could be projected near to each other in a random line.
 If the random line is chopped to buckets with length $w$, two points
 with smaller distance have higher probability to project into the
 same bucket.

\begin{equation}\label{eq:lsh_projection}
h_{a,b}(p) = \lfloor \frac{a \cdot p + b}{w}\rfloor
\end{equation}

Equation~\ref{eq:lsh_projection} shows the LSH function family for any data point $p \in \Re^d$.
$a$ is a $d$-dimensional random vector from Gaussian distribution, and $b$ is a real number
uniformly randomly drawn from $[0, w)$, where $w$ is the bucket width.  It is easy to get that 
two data points with distance $c$ has the probability to project into the same bucket is 
shown in Equation~\ref{bucket_probability}.

\begin{equation} \label{bucket_probability}
   Pr[c]    =   \int_{0}^{w}\frac{1}{c} f_{Y}(\frac{t}{c})   (1-\frac{t}{w})dt \\
\end{equation}

Since $a$ is drawn from Gaussian distribution which is a 2-stable distribution. 
For any vector $a$ from Gaussian distribution,  $a \cdot p_{i}-a \cdot p_{j} $ has the same distribution as 
 $\Vert p_{i}- p_{j}\Vert X  $, where $X$ follows Gaussian distribution \citep{zolotarev1986one}.

For any two data points $p_{i}$ and $p_{j}$ projected to the same bucket, there are two conditions.
The first one is  $|a \cdot p_{i}-a \cdot p_{j}| < w$, which is the projection distance less than the bucket width.
The second one is that no bucket boundary locates between the projection 
of $p_{i}$ and the projection of $p_{j}$ on the random line, 
and the probability is $1-\frac{|a \cdot p_{i}-a \cdot p_{j}| }{w}$.

\noindent For the first condition, 
\begin{equation}\label{eq:condition1}
\begin{aligned}
|a \cdot p_{i}-a \cdot p_{j} | < w &\Leftrightarrow | a \cdot (p_{i}-p_{j}) | < w  && \text{(2-stable distribution)}\\
            &\Leftrightarrow  | \Vert p_{i}- p_{j}\Vert X |< w \\
            &\Leftrightarrow    \Vert p_{i}- p_{j}\Vert | X |< w \\
            &\Leftrightarrow |X| < \frac{w}{\Vert p_{i}- p_{j}\Vert }  \\
\end{aligned}
\end{equation}

Since $X \sim N(0,1)$, then Y=|X| is a half normal distribution, whose probability density function is shown below
\begin{equation}\label{eq:half-normal-distribution}
   f_{Y}(y) = \frac{2}{\sqrt{2\pi}}e^{-\frac{y^2}{2}},  (y > 0) 
\end{equation}

According to Equation~\ref{eq:condition1}, the first condition is $Y <  \frac{w}{\Vert p_{i}- p_{j}\Vert }$. 
Let $c=\Vert p_{i}- p_{j}\Vert$, then

\begin{equation} 
\begin{aligned}
Pr[ Y < \frac{c}{w}] &= \int_{0}^{\frac{w}{c}} f_{Y}(y) dy  && \text{(let $t=cy$)} \\
                     &= \int_{0}^{w}  \frac{t}{c}f_{Y}(\frac{t}{c}) dt \\
                     &= \frac{1}{c} \int_{0}^{w}f_{Y}(\frac{t}{c}) dt \\
\end{aligned}
\end{equation}

\noindent For the second condition, let $Pr[c_{no}]$ denote the probability  that no bucket boundary 
locates between the projection 
of $p_{i}$ and the projection of $p_{j}$,

\begin{equation} 
\begin{aligned}
Pr[c_{no}]   &= 1-\frac{|a \cdot p_{i}-a \cdot p_{j}| }{w}] \\
                       &= 1-\frac{\Vert p_{i}-  p_{j}\Vert |X|}{w} \\
                       &=  1-\frac{cY}{w} \\
                       &=  1-\frac{t}{w} \\
\end{aligned}
\end{equation}

\begin{figure}[ht]
\centering
\includegraphics[width=0.7\textwidth]{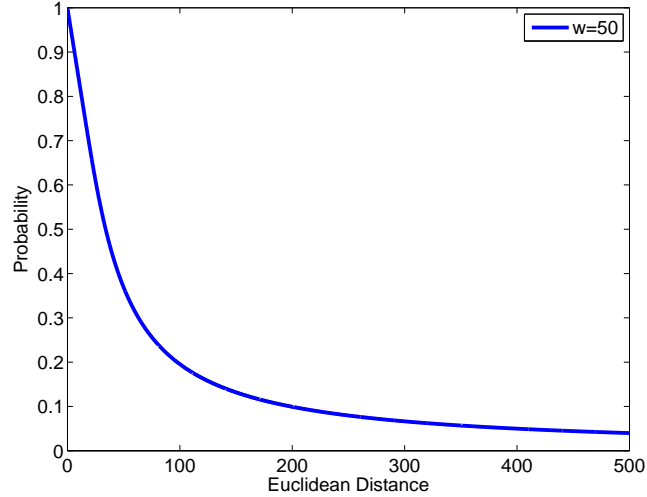}
\caption{Probability of two points hashed to the same bucket as a function of distance}
\label{fig:probability_same_buket}
\end{figure}

\noindent Therefore, for any data points $p_{i}$, $p_{j}$ and $c=\Vert p_{i}- p_{j}\Vert$, the probability that 
they are hashed to the same bucket is 
\begin{equation} 
\begin{aligned}
   Pr[c]   &= Pr[ Y < \frac{c}{w}]  Pr[c_{no}] \\
           &= \frac{1}{c} \int_{0}^{w}f_{Y}(\frac{t}{c}) dt  (1-\frac{t}{w}) \\
           &=   \int_{0}^{w}\frac{1}{c} f_{Y}(\frac{t}{c})   (1-\frac{t}{w})dt \\
\end{aligned}
\end{equation}
$Pr[c]$  is monotonically decreasing in $c$ when the bucket width is fixed, as shown in Figure~\ref{fig:probability_same_buket}.

In order to reduce the chance that two points far from each other hash to the same bucket,
normally, $K$ random lines are selected. If two data points are projected to the same
bucket in all $K$ random lines,  the two data points are stored in the same bucket in a 
hash table. Additionally, to increase the chance that two points near to each other
 hash to the same bucket, generally, $L$ hash tables are built on the data set.
Thus, for any data point $p$, the hash value $h(p)$ is shown as follow:

\begin{equation} 
h(p)=(h_{a_{1},b_{1}}(p), h_{a_{2},b_{2}}(p) \dots, h_{a_{K},b_{K}}(p)) \\
\end{equation}

To search neighbors for a data point $q$, all data points in the same hash bucket with $q$ in 
one of the hash tables will be validated.  The probability that two points with distance $c$
hash to the same bucket in of the of $L$ hash table is

\begin{equation}
Pr_{K,L}[c] = 1- (1- Pr[c]^K)^L
\end{equation}

\begin{figure}[ht]
\centering
\includegraphics[width=0.7\textwidth]{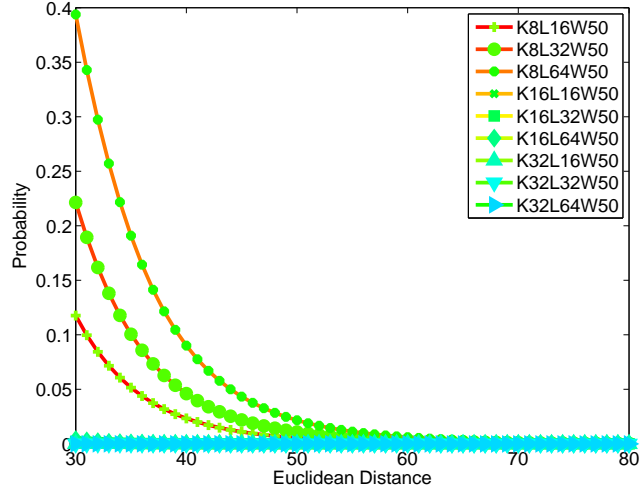}
\caption{Probability of two points hashed to the same bucket with different $K$ and $L$}
\label{fig:probability_diffKandL}
\end{figure}

Figure~\ref{fig:probability_diffKandL} shows several curves of probability that two points 
have the same hash value with different
$K$ and $L$. As we can see, the increasing of $K$ 
could decrease the probability, and the increasing of $L$ could increase the probability.


\section{Measurement of sequence clustering } \label{App:AppendixB}

\begin{equation}\label{eq:recall}
Recall=\frac{TP}{TP+FN}
\end{equation}

\begin{equation}\label{eq:precision}
Precision=\frac{TP}{TP+FP}
\end{equation}

\begin{equation}\label{eq:F1}
F_1=\frac{1}{\frac{2}{Recall}+\frac{1}{Precision}}=\frac{2\times Recall \times Precision}{Recall + Precision}
\end{equation}

\vspace{2em}
Normalized mutual information (NMI) is also used to assess clustering accuracy.
Let the set of clusters $G=\{g_{1}, g_{2}, g_{3},\dots,g_{s}\}$ is the ground truth,
and a set of clusters $C=\{c_{1}, c_{2}, c_{3},\dots,c_{t}\}$ is the clustering results from a  program.
Let $N$ denote the total number of protein sequences.
Then, NMI is defined as follow,
\begin{equation}\label{eq:NMI}
NMI(G,C)=\frac{2I(G,C)}{H(G) + H(C)} 
\end{equation}
where $I(G,C)$, $H(G)$ and $H(C)$ are defined  

\begin{equation}\label{eq:IGC}
I(G,C)=\sum_{i=1}^{s}\sum_{j=1}^{t} \frac{|g_{i}\cap c_{j}|}{N} log \frac{\frac{|g_{i}\cap c_{j}|}{N}}{\frac{|g_{i}|}{N}\frac{|c_{j}|}{N} }
\end{equation}
\begin{equation}\label{eq:HGC1}
H(G)=\sum_{i=1}^{s}\frac{|g_{i}|}{N}log \frac{|g_{i}|}{N} 
\end{equation}
\begin{equation}\label{eq:HGC2}
H(C)=\sum_{i=1}^{t}\frac{|c_{i}|}{N}log \frac{|c_{i}|}{N} 
\end{equation}

\end{document}